\documentclass[10pt]{iopart}

\usepackage{iopams}
\usepackage{bm}
\usepackage{cite}
\usepackage{mathrsfs}

\newcommand{\tfrac}{\textstyle\frac}

\begin{document}

\title{Quantum electrodynamical modes in pair plasmas} 

\author{Mattias Marklund, Padma K Shukla, Gert Brodin, and Lennart 
Stenflo}

\address{Department of Physics, Ume{\aa} University, SE--901 87
Ume{\aa}, Sweden}

\date{\today}

\begin{abstract}
We predict the existence of new nonlinear electromagnetic wave modes in pair plasmas. 
The plasma may be either non-magnetized or immersed in an external magnetic field.
The existence of these modes depends on the interaction of an intense circularly 
polarized electromagnetic wave with a plasma, where the nonlinear quantum vacuum 
effects are taken into account. This gives rise to new couplings between matter and radiation.  
We focus on pair plasmas, since the new modes are expected to exist in highly energetic 
environments, such as pulsar magnetospheres and the next generation of laser--plasma systems.   
\end{abstract}

%\maketitle

%\cite{Mihalas}
%\cite{Heisenberg-Euler}
%\cite{Bialynicka-Birula}
%\cite{Ding-Kaplan1,Ding-Kaplan2,Soljacic-Segev,%
%Brodin-marklund-Stenflo,Brodin-etal,Boillat}
%\cite{Ding-Kaplan2,Soljacic-Segev,Brodin-marklund-Stenflo} 
%\cite{Ding-Kaplan2},
%\cite{Soljacic-Segev}
%\cite{Brodin-marklund-Stenflo},
%\cite{Ding-Kaplan1},
%\cite{Brodin-etal}
%\cite{Boillat}
%\cite{Bialynicka-Birula,Boillat} 
%\cite{Thoma}
%\cite{Malomed-etal}
%\cite{Boomerang}

\section{Introduction.}

Quantum electrodynamics (QED) offers new phenomena with no classical 
counterparts, such as the Casimir effect. Similarly, and related to the 
Casimir effect, is so called photon--photon scattering (see, e.g., 
\cite{Heisenberg-Euler,Weisskopf,Schwinger,Greiner-Muller-Rafaelski}). 
The effective interaction between photons in a quantum vacuum is 
mediated by virtual electron--positron pairs, and therefore cannot occur 
within standard Maxwell electrodynamics. Photon--photon collisions have attracted 
much interest over the years, both from an experimental and an astrophysical 
point of view (see 
\cite{Bialynicka-Birula,Adler,Harding,Ding-Kaplan1,Latorre-Pascual-Tarrach,%
  Dicus-Kao-Repko,Ding-Kaplan2,Soljacic-Segev,Brodin-etal,%
  Brodin-marklund-Stenflo,Boillat,Heyl-Hernquist,%
  DeLorenci-Klippert-Novello,Thoma,Marklund-Brodin-Stenflo,Yu} 
and references therein). The effect of 
photon--photon scattering could be of fundamental importance in 
high-intensity laser pulses, in ultra-strong cavity fields, in the
surroundings of neutron stars and magnetars, and in the early Universe. 
However, the presence of plasmas in many highly energetic systems makes 
their theoretical analysis less tractable than the pure quantum vacuum model. 
Anyhow, we here present a theory of electromagnetic wave interaction in plasmas,
taking photon--photon scattering into account. It is shown that under certain 
circumstances the weak QED effects will act to generate distinct new wave 
modes. Specifically, we focus on pair plasmas, and argue that our new modes 
could be of importance in the next generation of laser--plasmas, as well as in 
pulsar magnetospheres.

\section{Basic equations.}

The weak nonlinear self-interaction of photons in the quantum 
vacuum can be expressed in terms of the Heisenberg--Euler Lagrangian 
\cite{Heisenberg-Euler} 
\begin{equation}\label{eq:lagrangian}
  \mathscr{L} = \epsilon_0\mathscr{F} +
  \kappa\epsilon_0^2\left(4\mathscr{F}^2 + 7\mathscr{G}^2 \right),
\end{equation}
where $\mathscr{F} = -F_{ab}F^{ab}/4 = (E^2 - c^2B^2)/2$, 
$\mathscr{G} = -F_{ab}\widehat{F}^{ab}/4 = c\bm{E}\cdot\bm{B}$, and 
$\widehat{F}_{ab} = \epsilon_{abcd}F^{cd}/2$ is the dual of Maxwell's field 
strength tensor $F_{ab}$.
Here $\kappa \equiv 2\alpha^2\hbar^3/45m_e^4c^5 \approx 1.63\times
10^{-30}\, \mathrm{m}\mathrm{s}^{2}/\mathrm{kg}$, $\alpha$ is
the fine-structure constant, $\hbar$ is the Planck constant divided by 
$2\pi$, $m_e$ is the electron mass, and $c$ is the speed of light in vacuum. 
The Lagrangian (\ref{eq:lagrangian}) is valid when 
\begin{equation}
  \omega \ll \omega_e \equiv m_ec^2/\hbar ,  \qquad |\bm{E}| \ll
  E_{S} \equiv 
  m_ec^2/e\lambda_c   
  \label{eq:constraint} 
\end{equation}
respectively. Here $e$ is the elementary charge, $\lambda_c$ the
Compton wavelength, $\omega_e$ the Compton frequency, and $E_{S} \simeq
10^{18}\,\mathrm{V}/\mathrm{m}$ the Schwinger field strength. 
The first inequalility states that the individual photons should not create 
real electron--positron pairs out of vacuum fluctuations, while the second 
states that the collective energy of many photons should not create
real electron--positron pairs. 

Using Eq.\ (\ref{eq:lagrangian}), the dispersion relation, in the absence of matter fields, 
for photons in a background electromagnetic field
$\bm{E}$, $\bm{B}$ is \cite{Bialynicka-Birula,Boillat}
\begin{equation}\label{eq:dispersionrelation}
  \omega(\bm{k}, \bm{E}, \bm{B}) = c|\bm{k}|\left( 1 -
  \case{1}{2}\lambda|\bm{Q}|^2 \right) ,
\end{equation}
where
\begin{eqnarray}
  |\bm{Q}|^2 = \varepsilon_0\left[ E^2 + c^2B^2 - 
   (\hat{\bm{k}}\cdot\bm{E})^2 -
   c^2(\hat{\bm{k}}\cdot\bm{B})^2 -
   2c\hat{\bm{k}}\cdot(\bm{E}\times\bm{B})\right] ,
\label{eq:Q2}
\end{eqnarray}
and $\lambda = \lambda_{\pm}$, where $\lambda_+ = 14\kappa$ and
$\lambda_- = 8\kappa$ for the two different polarisation states of the
photon. Furthermore, $\hat{\bm{k}} \equiv \bm{k}/k$.

We may add the matter fields to the Lagrangian (\ref{eq:lagrangian}). Introducing the vector 
potential $A^b$, such that $F_{ab} = \partial_aA_b - \partial_bA_a$, Euler--Lagrange's equations 
give us the sourced Maxwell equations
\begin{equation}  \label{eq:maxwell}
  \partial_bF^{ab} = 2\epsilon_0\kappa\partial_b\left[ 
  (F_{cd}F^{cd})F^{ab} 
  + \tfrac{7}{4}(F_{cd}\widehat{F}^{cd})\widehat{F}^{ab} \right]
  + \mu_0 j^a ,
\end{equation}
where $j^a$ is the four current. Using the Lorentz gauge $\partial_bA^b = 0$, Eq.\ 
(\ref{eq:maxwell}) yields
\begin{equation}\label{eq:maxwell2}
\fl  \left[ 1 - 2\epsilon_0\kappa(F_{cd}F^{cd})\right]\Box A^a 
  = 2\epsilon_0\kappa\left[ F^{ab}\partial_b(F_{cd}F^{cd}) 
  + \tfrac{7}{4}\widehat{F}^{ab}\partial_b(F_{cd}\widehat{F}^{cd}) \right] 
  + \mu_0j^a ,
\end{equation}
where $\Box = \partial_a\partial^a$ is the d'Alambertian. 

For a circularly polarized electromagnetic wave $\bm{E}_0 
= E_0(\hat{\bm{x}} \pm i\hat{\bm{y}})\exp(i\bm{k}\cdot\bm{r} - i\omega t)$ 
propagating along a constant magnetic field $\bm{B}_0 = B_0\hat{\bm{z}}$, 
the invariants satisfy
\begin{equation} 
  F_{cd}F^{cd} = -2E_0^2\left( 1 - \frac{k^2c^2}{\omega^2}\right) 
    + 2c^2B_0^2 , 
  \qquad 
  F_{cd}\widehat{F}^{cd} = 0 ,
\end{equation}
where $\bm{k}$ is the wave vector and $\omega$ the frequency of the
circularly polarized electromagnetic wave. Using these expressions,
Eq.\ (\ref{eq:maxwell2}) reduces to
\begin{equation}  \label{eq:wave}
  \Box A^a = -4\epsilon_0\kappa\left[ E_0^2\left( 1 - \frac{k^2c^2}{\omega^2}\right) 
    - c^2B_0^2 \right]\Box A^a + \mu_0 j^a ,
\end{equation}
which, together with the dynamical equations for the particle current $j^a$, is our 
main equation.

\section{Unmagnetized plasmas}

With $B_0 = 0$, the effect due to the presence of a plasma can be written in terms of a 
modified wave operator 
\begin{equation}
  \Box \rightarrow \Box - \frac{\omega _{p}^{2}}{c^{2}} , \label{eq:substitution}
\end{equation}
where the plasma frequency is given by \cite{Stenflo1976,Stenflo-Tsintsadze1979}
\begin{equation}\label{eq:plasma}
  \omega_p = \sum_j \gamma_j^{-1/2} \omega_{pj} 
  = \sum_j \left( \frac{n_{0j}q_{j}^{2}}{\epsilon_{0}m_{j}\gamma_{j}} \right)^{1/2} ,
\end{equation}
where the sum is over particle species, $q_j$ is the charge, $m_j$ is the rest mass, $n_{0j}$ denotes the particle
density in the laboratory frame, and the relativistic factor of each
particle species is  
$\gamma_j = (1 + q_j^2 E_0^2/m_j^2 c^2 \omega^2)^{1/2}$.
%where $E_0$ denotes the absolute value of the electric field amplitude.
Making a harmonic decomposition of the fields, we see that Eq.\ (\ref{eq:substitution}) 
gives
\begin{equation}\label{eq:harmonic}
  \Box  - \frac{\omega_p^2}{c^2} = \frac{\omega^2 
  - \omega_p^2}{c^2} - k^2.
\end{equation}

In the low frequency limit, $\omega^2 \ll k^2 c^2$,  we 
obtain from (\ref{eq:maxwell2}) and ({\ref{eq:harmonic}) the nonlinear dispersion relation 
\begin{equation}
\omega^2 = \frac{2\alpha}{45\pi}\left(
  \frac{E_0}{E_{S}} \right)^2 \frac{k^4 c^4}{\omega_p^2+k^2
  c^2} .
  \label{Dispersion-relation} 
\end{equation}
Next we focus our attention on a pair plasma. 
For an equal density ($n_0$) electron--positron plasma, with ultra-relativistic particle motion 
($\gamma_e \gg 1$), we use the approximation
$\omega_p^2 \approx 2\omega_{pe}^2
(\omega/\omega_e) (E_{S}/E_0)$, where $\omega_e = m_ec^2/\hbar$ and 
$\omega_{pe} = (e^2 n_0/\epsilon_0 m_e)^{1/2}$. 
We then obtain \cite{Stenflo-Brodin-Marklund-Shukla}  
\begin{equation}
\omega^3 = \frac{\alpha}{45\pi}\left( \frac{\omega_e}{
  \omega_{pe}} \right) \left(
  \frac{E_0}{E_{S}} \right)^3\frac{k^4 c^4}{\omega_{pe} + 
  (E_0/E_{S})(kc\omega_e /2\omega\omega_{pe})kc} .
\label{Relativistic-DR}
\end{equation}
from Eq.\ (\ref{Dispersion-relation}).
We note that the Compton frequency $\omega_e$ is much larger than $\omega_{pe}$ for  
virtually all plasmas, and corresponds to electron densities up to $\sim 10^{38}$ 
m$^{-3}$.

If the amplitude of the vector potential varies slowly, we can derive a 
nonlinear Schr\"odinger equation by taking the media response into account. We may take 
the scalar potential $\phi = 0$. The weakly varying vector potential amplitude $\bm{A} 
= \bm{A}(t,z)$ then satisfies \cite{Hasegawa} 
\begin{equation}\label{eq:nlse}
  i\left( \frac{\partial}{\partial t} + v_g\frac{\partial}{\partial z} \right)\bm{A}
  + \frac{v^{\prime}_g}{2}\frac{\partial^2\bm{A}}{\partial z^2} 
  + a(|\bm{A}|^2 - A_0^2)\bm{A}  
  = 0 ,
\end{equation}
where $v_g = \partial\omega/\partial k$, $v^{\prime}_g = \partial^2\omega/\partial k^2$,
$a = -\partial \omega/\partial A_0^2$, and $A_0^2 = E_0^2/\omega^2$. We note that 
the response from the medium through $\omega_{pe}$ depends on several parameters.

\section{Magnetized plasmas}

In the presence of a magnetic field, the plasma contribution to the dispersion relation can be 
obtained by substituting
\begin{equation}
  \Box \rightarrow \Box 
  - \sum_j\frac{\omega\omega_{pj}^2}{\omega\gamma_j \pm \omega_{cj}} ,
\label{eq:dispersionfunction}
\end{equation}
for the d'Alembertian. The sum is over the plasma particle species $j$,  
\begin{equation}
  \omega_{cj} = \frac{q_jB_0}{m_{j}} , 
\label{eq:gyro}
\end{equation}
is the gyrofrequency, and 
\begin{equation}
  \gamma_j = (1 + \nu^{2}_j)^{1/2}, 
\end{equation}
is the the gamma factor of species $j$, with $\nu_j$ satisfying \cite{Stenflo1976,Stenflo-Tsintsadze1979}
\begin{equation}
  \nu^{2}_j = \left(
  \frac{eE_0}{cm_{j}} \right)^2\frac{1 + \nu^{2}_j}{[\omega(1 +
  \nu^{2}_j)^{1/2} \pm \omega_{cj}]^2} .
  \label{eq:nu}
\end{equation}

Making a harmonic decomposition of the fields, and looking for 
low-frequency modes in an ultra-relativistic pair plasma, we use the 
approximations $\omega \ll kc$ and $\gamma_e \gg 1$, at which Eq.\ 
(\ref{eq:maxwell2}) together with (\ref{eq:dispersionfunction}) gives
\cite{Marklund-Shukla-Stenflo-Brodin}
\begin{equation}
   \frac{k^{2} c^{2} }{\omega^{2} } \approx \frac{4\alpha}{45\pi}\left[\left( 
    \frac{E_0}{E_S}  \right)^2 \frac{k^{2} c^{2} }{\omega^{2} } 
    + \left(\frac{cB_0}{E_S}\right)^2 \right]\frac{k^2c^2}{\omega^2}  \mp 
    \frac{\omega_{pe}^{2}}{\omega\omega_e}\frac{E_S}{E_0}  .  
\label{eq:transverse3}
\end{equation}
In the limit of no photon--photon scattering, i.e.\ $\alpha \rightarrow 0$, we recover 
the modes found in Ref.\ \cite{Stenflo-Tsintsadze1979}.

Magnetized pair plasmas can be found in the surroundings of pulsars and strongly magnetized stars, 
e.g.\ in the form of accretion disks. At a distance from the star's surface, the magnetic field will be 
weak, being essentially dipole in character, and the first term in the square bracket of 
Eq.\ (\ref{eq:transverse3}) will be the dominant QED contribution.   

Close to neutron stars or magnetars, the magnetic field strengths are in the range 
$10^6 - 10^{11}\, \mathrm{T}$ \cite{magnetosphere,magnetar}, and, depending on the frequency of 
the circularly polarized wave, the second term in the square bracket of Eq.\ (\ref{eq:transverse3}) 
may dominate the behavior of the wave mode. If $cB_0 \gtrsim E_S$, we have
\begin{equation}
  \omega \approx \mp \omega_e
  \frac{E_0}{E_S}\left[ 1 - \frac{4\alpha}{45\pi}
  \left(\frac{cB_0}{E_S}\right)^2 \right]
  \left(\frac{kc}{\omega_{pe}}\right)^2   .
\end{equation}

\section{Discussion and conclusion}

Most situations in which photon--photon scattering can be important are of an extreme nature. 
Examples of environments where the effects may either be dynamically significant, or measurable, are 
the next generation of high power lasers and possibly their combination with plasmas into 
laser--plasma systems \cite{Soljacic-Segev,Yu}, high field superconducting 
cavities \cite{Brodin-marklund-Stenflo}, and 
astrophysical environments, such as pulsar magnetospheres \cite{magnetosphere} and the 
vicinity of magnetars \cite{magnetar}. In astrophysical environments, effects such as photon splitting or
magnetic lensing have been suggested to take place \cite{Bialynicka-Birula,Adler,Harding,Heyl-Hernquist}. 
Even in cosmology, the effects of photon--photon scattering could be detectable using precision observations 
of the cosmic microwave background \cite{wmap,Marklund-Brodin-Stenflo}.

However, plasmas may in many circumstances be a prominent component of the physical systems considered above. 
For example, it is currently believed that the highest experimental field strengths could be obtained using 
laser--plasma systems \cite{bob}. Therefore, the addition of plasmas to the dynamics of photon--photon 
scattering adds an important piece to our understanding of the nonlinear quantum vacuum, and as shown here, 
could provide a unique signature of photon--photon scattering. It remains to be seen whether this can 
be realized in a laboratory or in astrophysics.

%\newpage


\section*{References}

\begin{thebibliography}{99}

%  \bibitem{Mihalas} D.\ Mihalas and B.\ Weibel--Mihalas,
%  \textit{Foundation of Radiation Hydrodynamics} (Dover Publications,
%  1999).

  \bibitem{Heisenberg-Euler} W.\ Heisenberg and H.\ Euler, Z.\
  Phys.~\textbf{98} 714 (1936).
  
  \bibitem{Weisskopf} 
  V.S.\ Weisskopf, K.\ Dan.\
  Vidensk.\ Selsk.\ Mat.\ Fy.\ Medd.~\textbf{14} 1 (1936). 
  
  \bibitem{Schwinger}
  J.\ Schwinger, Phys.\ Rev.~\textbf{82}
  664 (1951).
  
  \bibitem{Greiner-Muller-Rafaelski}
  W.\ Greiner, B.M\"uller and J.\ Rafelski,
  \textit{Quantum electrodynamics of strong fields} (Springer, Berlin,
  1985).

  \bibitem{Bialynicka-Birula} Z.\ Bialynicka--Birula and I.\
  Bialynicki--Birula, Phys.\ Rev.\ D \textbf{2} 2341 (1970).
  
  \bibitem{Adler} S.L.\ Adler, Ann.\ Phys.-NY  \textbf{67} 599 (1971).  

  \bibitem{Harding} A.K.\ Harding, Science \textbf{251} 1033 (1991).

  \bibitem{Ding-Kaplan1}
  A.E.\ Kaplan and Y.J.\ Ding,
    Phys.\ Rev.\ A \textbf{62} 043805 (2000).
  
  \bibitem{Latorre-Pascual-Tarrach}
  J.I.\ Latorre, P.\ Pascual and R.\ Tarrach,
    Nucl.\ Phys.\ B \textbf{437} 60 (1995).
  
  \bibitem{Dicus-Kao-Repko}
  D.A.\ Dicus, C.\ Kao and W.W.\ Repko,
    Phys.\ Rev.\ D \textbf{57} 2443 (1998).

  \bibitem{Ding-Kaplan2}
  Y.J.\ Ding and A.E.\ Kaplan,
    Phys.\ Rev.\ Lett.~\textbf{63} 2725 (1989).

  \bibitem{Soljacic-Segev} M.\ Solja\v{c}i\'c and M.\ Segev,
    Phys.\ Rev.\ A \textbf{62} 043817 (2000).

  \bibitem{Brodin-etal} G.\ Brodin, L.\ Stenflo, D.\ Anderson, M.\
   Lisak, M.\ Marklund and P.\ Johannisson, Phys.\ Lett. A
  \textbf{306} 206 (2003).

  \bibitem{Brodin-marklund-Stenflo} G.\ Brodin, M.\ Marklund and L.\
  Stenflo, Phys.\ Rev.\ Lett.~\textbf{87} 171801 (2001).

  \bibitem{Boillat} G.\ Boillat, J.\ Math.\ Phys.~\textbf{11} 941
  (1970).
  
  \bibitem{Heyl-Hernquist}
  J.S.\ Heyl and L.\ Hernquist, J.\ Phys.\ A: Math.\ Gen.~\textbf{30}
  6485 (1997).
  
  \bibitem{DeLorenci-Klippert-Novello}
  V.A.\ De Lorenci, R.\ Klippert, M.\
  Novello and J.M.\ Salim, Phys.\ Lett.\ B \textbf{482} 134
  (2000).

  \bibitem{Thoma} M.H.\ Thoma, Europhys.\ Lett.~\textbf{52}
  498 (2000).
  
  \bibitem{Marklund-Brodin-Stenflo} M.\ Marklund, G.\ Brodin and L.\
  Stenflo, Phys.\ Rev.\ Lett.\ \textbf{91} 163601 (2003).
  
  \bibitem{Yu} B.\ Shen and M.Y.\ Yu, Phys.\ Plasmas \textbf{10} 4570
  (2003). 
  
  \bibitem{Stenflo1976} L.\ Stenflo, Phys.\ Scripta {\bf 14}, 320 (1976).

\bibitem{Stenflo-Tsintsadze1979} L.\ Stenflo and N.L.\ Tsintsadze, 
  Astrophys. Space Sci.\ \textbf{64}, 513 (1979).
  
  \bibitem{Stenflo-Brodin-Marklund-Shukla} L.\ Stenflo, G.\ Brodin,
    M.\ Marklund, and P.K.\ Shukla, arXiv physics/0410090 (2004).
 
  \bibitem{Hasegawa} A.\ Hasegawa, \textit{Plasma Instabilities and
  Nonlinear Effects} (Springer-Verlag, Berlin, 1975).
     
  \bibitem{Marklund-Shukla-Stenflo-Brodin} M.\ Marklund, P.K.\ Shukla,
    L.\ Stenflo, and G.\ Brodin, arXiv astro-ph/0410294 (2004). 
  
%  \bibitem{Malomed-etal} B. Malomed \textit{et al.},
%    Phys.\ Rev.\ E \textbf{55} 962 (1997).
 
%  \bibitem{Boomerang} P.\ de Bernardis \textit{et al.}, Nature
%  \textbf{404} 955 (2000). 
 
%  \bibitem{Karpman1} V.I.\ Karpman, Plasma Phys.\ \textbf{13} 477
%  (1971).
  
%  \bibitem{Zakharov} V.E.\ Zakharov, Sov.\ Phys.-JETP \textbf{35}
%    908 (1972).

%  \bibitem{Karpman2} 
%  V.I.\ Karpman, Phys.\ Scr.\ \textbf{11} 263 (1975).

%  \bibitem{Karpman3}
%  V.I.\ Karpman, \textit{Nonlinear Waves in Dispersive Media}
%  (Pergamon Press, Oxford, 1975).

%  \bibitem{Shukla} P.K.\ Shukla, Phys.\ Scr.\ \textbf{45} 618 (1992).

%  \bibitem{Mendonca} J.T.\ Mendon\c{c}a, \textit{Theory of Photon
%  Acceleration} (Institute of Physics Publishing, Bristol, 2001).

%  \bibitem{Desaix-Anderson-Lisak} M.\ Desaix, D.\ Anderson and M.\
%  Lisak, J.\ Opt.\ Soc.\ Am.\ B \textbf{8} 2082 (1991).
  
%  \bibitem{collapse} E.A.\ Kuznetsov, A.M.\ Rubenchik and V.E.\ Zakharov,
%  Phys.\ Rep.\ \textbf{142} 103 (1986); J.J.\ Rasmussen and K.\
%  Rypdal, Physica Scripta \textbf{33} 481 (1986).
  
%  \bibitem{Mamaev-etal} S.G.\ Mamaev, V.M.\ Mostepanenko and M.I.\
%  E\u{\i}des, Sov.\ J.\ Nucl.\ Phys.\ \textbf{33} 569 (1981). 

%  \bibitem{Piran} T.\ Piran, Phys.\ Rep.~\textbf{314} 575 (1999).

\bibitem{magnetosphere} V.I.\ Beskin {\it et al.}, 
  \textit{Physics of the Pulsar Magnetosphere}
    (Cambridge: Cambridge University Press, 1993).
  
  \bibitem{magnetar} C.\ Kouveliotou, S.\ Dieters, T.\ Strohmayer
  \textit{et al.}, Nature \textbf{393} 235 (1998). 
%  Note that the surface field $10^{10}\,
%  \mathrm{T}$ is close to the critical value $E_{\text{crit}}/c$, but
%  because of the dipole nature of the magnetic field, the field
%  strength decreases rapidly from the surface and out, and thus  
%  the second of the inequalities (\ref{eq:constraint}) is also
%  satisfied. 
  
%  \bibitem{Kondratyev} V.N.\ Kondratyev, Phys.\ Rev.\
%  Lett.~\textbf{88} 221101 (2002).

%  \bibitem{Gorbunov-etal} L.M.\ Gorbunov and V.I.\ Kirsanov, Zh.\
%    Eksp.\ Teor.\ Fiz.\ \textbf{93} 509 (1987) [Sov.\ Phys.-JETP
%    \textbf{66} 290 (1987)].
  
  \bibitem{wmap} URL $\mathtt{http:\!//map.gsfc.nasa.gov/m\_mm.html}$
  
  \bibitem{bob} R.\ Bingham, Nature \textbf{424} 258 (2003).
   
\end{thebibliography}
\end{document}